\documentclass[aps,pra,twocolumn,groupedaddress,nofootinbib,showpacs]{revtex4-1}
\usepackage{amsmath}
\usepackage{amssymb}
\usepackage{mathrsfs}
\usepackage[pdftex,colorlinks=true,urlcolor=blue,linkcolor=blue,citecolor=blue]{hyperref}
\usepackage{graphicx}
\usepackage{float}
\usepackage[abs]{overpic}
\usepackage{color}
\usepackage{subfigure}
\begin{document}

\title{Dimensional Crossover in a Spin-imbalanced Fermi gas}

\author{Shovan Dutta}
%\email{sd632@cornell.edu}
\author{Erich J. Mueller}
%\email{em256@cornell.edu}
\affiliation{Laboratory of Atomic and Solid State Physics, Cornell University, Ithaca, New York 14850, USA}

\date{\today}

\begin{abstract}
We model the one-dimension (1D) to three-dimension (3D) crossover in a cylindrically trapped Fermi gas with attractive interactions and spin-imbalance. We calculate the mean-field phase diagram, and study the relative stability of exotic superfluid phases as a function of interaction strength and temperature. For weak interactions and low density, we find 1D-like behavior, which repeats as a function of the chemical potential as new channels open. For strong interactions, mixing of single-particle levels gives 3D-like behavior at all densities. Furthermore, we map the system to an effective 1D model, finding significant density dependence of the effective 1D scattering length.
\end{abstract}

\pacs{67.85.Lm, 74.20.-z, 03.75.Hh, 71.10.Pm}

\maketitle

\section{\label{introduction}Introduction}
Spin-imbalanced Fermi gases are predicted to display an array of exotic superconducting phases, where the order parameter has non-trivial structure \cite{3DsmallFFLO+BP1indeepBEC+detection+noncond, 3DsmallFFLO+BPintro+BP1indeepBEC+detection, 3DsmallFFLO+BPintro+BP1indeepBEC, 3DsmallFFLO+BP1indeepBEC+noncond, 3DsmallFFLO+BP1indeepBEC, 3DsmallFFLO, 3DsmallFFLO+Hartree+regularizationLDA, 3D+ASLDA, 1DrichphasesMF, 1DrichphasesMF+mapping+1Dphasediagram+exactvsMF, 1Drichphases+1Dexperiment+1Dphasediagram, 1Drichphases+1Dphasediagram+mapping, 1Drichphases+1Dphasediagram, 1Drichphases, 1Drichphases+quasi1Dpromising, 1drichphases+detection, FR+, DFS+detection, mixedphase+detection, BP+display, detection+, noncond+, display, liquidcrystal}. Mean-field theories predict that these states occupy a very small fraction of the phase diagram in 3D, but are ubiquitous in 1D \cite{3DsmallFFLO+BP1indeepBEC+detection+noncond, 3DsmallFFLO+BPintro+BP1indeepBEC+detection, 3DsmallFFLO+BPintro+BP1indeepBEC, 3DsmallFFLO+BP1indeepBEC+noncond, 3DsmallFFLO+BP1indeepBEC+noncond, 3DsmallFFLO+BP1indeepBEC, 3DsmallFFLO, 3DsmallFFLO+Hartree+regularizationLDA, 1DrichphasesMF, 1DrichphasesMF+mapping+1Dphasediagram+exactvsMF}, with the caveat that quantum fluctuations prevent long-range order in 1D \cite{longrangeorder}. Indeed, cold-atom experiments in 3D \cite{3dexperiments} have found no sign of the exotic Fulde-Ferrell-Larkin-Ovchinnikov (FFLO) phase \cite{FFLO}, while experiments on 1D tubes \cite{1Drichphases+1Dexperiment+1Dphasediagram} found thermodynamic evidence for a fluctuating version \cite{1Drichphases+1Dphasediagram+mapping, 1Drichphases+1Dphasediagram, 1Drichphases, 1Drichphases+quasi1Dpromising, 1drichphases+detection} of FFLO, but were unable to measure the order parameter. One avenue for directly observing these exotic superfluid states is to use highly anisotropic quasi-1D geometries where they should be robust \cite{1Drichphases+quasi1Dpromising, quasi1Dpromising+quasi1Dsingleband+mapping, quasi1Dpromising+detection, quasi1Dfinitesize+BP, quasi1Dfinitesize, quasi1Dpromising, quasi1Dpromising+quasi1Dsingleband+detection, quasi1Dpromising+bolechFFLOdetection, quasi1Dpromising+quasi1Dsingleband, quasi1Dpromising+quasi1Dfinitesize+ASLDA}. Here we solve the Bogoliubov-de-Gennes (BdG) equations in such a geometry. We find large regions of the phase diagram in which the FFLO finite-momentum-pairing state is stable. We also find a stable breached-pair (BP) state where pairs coexist with a Fermi surface \cite{3DsmallFFLO+BPintro+BP1indeepBEC+detection, 3DsmallFFLO+BPintro+BP1indeepBEC, BPintro+detection, BPintro, BPintro+interiorgap, BPintro+quarkmatter+BP1indeepBEC+topology}. Our analysis provides a much needed narrative for thinking about the 1D-3D crossover, going beyond the existing single-band models \cite{quasi1Dpromising+quasi1Dsingleband+mapping, quasi1Dsingleband+mapping, quasi1Dsingleband, quasi1Dpromising+quasi1Dsingleband+detection, quasi1Dpromising+quasi1Dsingleband} and studies of finite systems \cite{quasi1Dfinitesize+BP, quasi1Dfinitesize, quasi1Dpromising+quasi1Dfinitesize+ASLDA}. While we focus on cold atoms, these considerations are also relevant to nuclear, astrophysical \cite{condmatter+quarkmatter, BPintro+quarkmatter+BP1indeepBEC+topology, quarkmatter}, and condensed-matter systems \cite{condmatter+quarkmatter, condmatter+FFLOdetected, condmatter}. Evidence  of the FFLO phase has recently been found in a quasi-2D superconductor \cite{condmatter+FFLOdetected}, and there are ongoing attempts to see related physics in 2D atomic systems \cite{2DatomsFFLO}.

We consider a harmonic oscillator potential of frequency $\omega_{\perp}$ which confines the motion of the atoms in the $x$-$y$ plane. The atoms are free to move in the $z$ direction, have mass $m$, and interact via $s$-wave collisions, characterized by a scattering length $a_s$ (tuned via a Feshbach resonance \cite{FR+, FRreview}). We consider the ``Bardeen-Cooper-Schrieffer (BCS) side" of resonance where $a_s < 0$, and calculate the mean-field phase diagram in the $\mu$-$h$ plane, where $\mu \equiv (\mu_{\uparrow} + \mu_{\downarrow})/2$ and $h \equiv (\mu_{\uparrow} - \mu_{\downarrow})/2$ denote respectively the average chemical potential and the chemical potential difference of the two spins. Prior work on this model has examined the low-density (small $\mu$) limit, where the transverse motion of the atoms are confined to the lowest oscillator level \cite{1DrichphasesMF+mapping+1Dphasediagram+exactvsMF, 1Drichphases+1Dexperiment+1Dphasediagram, 1Drichphases+1Dphasediagram+mapping, quasi1Dpromising+quasi1Dsingleband+mapping, quasi1Dsingleband+mapping}, and the system maps onto an effective 1D model \cite{mapping}. Conversely, when $\mu$ is large, the atoms can access many energy levels of the trap, and the system is locally three-dimensional. Here we investigate the crossover between these regimes.

The exact 1D phase diagram contains three phases which are fluctuating analogs of the BCS superfluid, the FFLO state, and a fully polarized (FP) gas \cite{1DrichphasesMF+mapping+1Dphasediagram+exactvsMF, 1Drichphases+1Dexperiment+1Dphasediagram, 1Drichphases+1Dphasediagram+mapping, 1Drichphases+1Dphasediagram}. Since interaction effects in 1D are stronger at low densities, pairs are more stable at smaller $\mu$, and the slope $\gamma \equiv d\mu/dh$ of the line separating the BCS and FFLO phases has a negative slope. The analogous phase boundary in 3D has a positive slope, providing a convenient distinction between 1D-like and 3D-like behavior. In 3D there is also a partially polarized Normal (N) state \cite{3DsmallFFLO+BPintro+BP1indeepBEC+detection, 3DsmallFFLO+BPintro+BP1indeepBEC}.

\begin{figure}[t]
\includegraphics[width=\columnwidth]{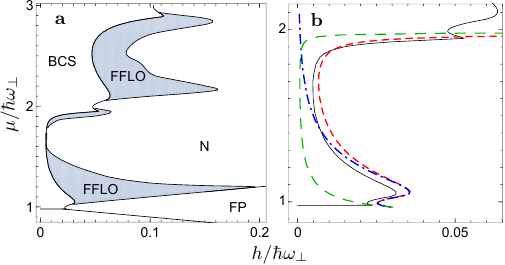}
\caption{\label{phase_diagram_weak_interaction}(Color online) Zero-temperature phase diagram of a two-component Fermi gas in a 2D harmonic trap of frequency $\omega_{\perp}$. Here $d_{\perp}/a_s = -3$, where $a_s$ is the 3D scattering length, and $d_{\perp} \equiv$ $(\hbar/m\omega_{\perp})^{1/2}$. {\bf a}: Phase boundaries calculated using 3D BdG equations. {\bf b}: BCS critical field of the full model (solid curve) and of various effective 1D models (dashed curves). Short-dashed (red): 1D BdG with the mapping in Eq. (\ref{mapping}), dot-dashed (blue): 1D BdG with Olshanii's mapping \cite{mapping}, long-dashed (green): Bethe ansatz with Eq. (\ref{mapping}).}
\end{figure}

For weak interactions and $\mu < 2\hbar \omega_{\perp}$, we find 1D-like behavior, in that $\gamma < 0$. The critical field jumps whenever a new channel opens (near $\mu \sim n \hbar \omega_{\perp}$), but after this jump we again find $\gamma < 0$ (Fig. \ref{phase_diagram_weak_interaction}a). Once many channels are occupied we find 3D-like behavior with $\gamma>0$. Each 1D-like interval hosts a large FFLO region. In the 3D regime, these regions merge to form a single domain. As interactions are increased, the crossover to 3D-like behavior moves to smaller $\mu$ (Fig. \ref{phase_diagram_stronger_interactions}a). For very strong interactions near unitarity ($a_s \to -\infty$), the harmonic oscillator levels are strongly mixed, and we always find $\gamma>0$. (Fig. \ref{phase_diagram_stronger_interactions}b). Regardless, we find that the FFLO phase occupies much of the phase diagram for all interaction strengths. Moreover, at sufficiently strong interactions, we find a BP region, nestled between the BCS and the FFLO phases. Such a (zero-temperature) BP phase is stable in 3D only for negative $\mu$ in the deep Bose-Einstein condensate (BEC) side of resonance ($a_s > 0$) \cite{3DsmallFFLO+BP1indeepBEC+detection+noncond, 3DsmallFFLO+BPintro+BP1indeepBEC+detection, 3DsmallFFLO+BPintro+BP1indeepBEC, 3DsmallFFLO+BP1indeepBEC+noncond, 3DsmallFFLO+BP1indeepBEC, BPintro+quarkmatter+BP1indeepBEC+topology, BP1indeepBEC}. These results clearly show that exotic superfluids will be observable in quasi-1D experiments.

We study the temperature variation of the phase diagrams (Figs. \ref{temperature_variation1} and \ref{temperature_variation2}). The FFLO and BP phases shrink much faster with temperature than the BCS phase as they have much smaller pairing energies. For weak interactions, the BCS phase survives in isolated pockets, which disappear sequentially with temperature. The critical temperatures grow with interactions, as interactions favor pairing.

In addition to directly solving the 3D BdG equations, we map the system to an effective 1D model in the single-channel limit $\mu < 2 \hbar \omega_{\perp}$. We find that the effective 1D coupling constant $g_{\mbox{\scriptsize{1D}}}$ becomes more strongly attractive at larger $\mu$. Our mapping reduces to that in \cite{mapping} in the low-density limit, but has previously unexplored correction terms at higher densities. These become more important at stronger interactions (Eq. (\ref{mapping})).

We use the Bogoliubov-de-Gennes (BdG) mean-field formalism. This approach does not include a Hartree self-energy \cite{3DsmallFFLO+Hartree+regularizationLDA}. This deficiency is typically unimportant for weak interactions, but becomes significant as one approaches unitarity. It may also be important for studying the competition between phases with similar energies. Unfortunately the literature contains no convenient way to incorporate the Hartree term. The technical difficulty is that the bare coupling constant for contact interactions has an ultraviolet divergence. Renormalizing this divergence causes the Hartree term to identically vanish, and there is active debate about the significance of those terms \cite{3DsmallFFLO+Hartree+regularizationLDA}. At unitarity, one can circumvent this problem by imposing universality on the equation of state, and constructing a regularized energy functional \cite{quasi1Dpromising+quasi1Dfinitesize+ASLDA, 3D+ASLDA, DFT}. However, there is no equivalent scheme at intermediate interactions, as the proper set of constraints are unknown. Along with self-energy corrections, quantum fluctuations also become significant at stronger interactions \cite{fluctuations+noncond}. Thus we do not expect our results to be accurate in the unitary regime. In fact, a recent experiment with ${}^6$Li atoms, performed near unitarity, found the behavior of the system to be 1D-like at low densities \cite{1Drichphases+1Dexperiment+1Dphasediagram}, whereas our model predicts 3D-like physics there. We believe the physics neglected in the BdG approach largely renormalizes $a_s$, and that our unitary results should agree with experiments for $a_s > 0$. Lastly, we cannot rule out other phases not considered here, e.g., a state with deformed Fermi surface pairing \cite{DFS+detection}, or an incoherent mixture of paired and unpaired fermions \cite{mixedphase+detection}.

Despite these limitations, our simple model lets us make concrete predictions, and provides insight into the nature of the dimensional crossover. In particular, we find that the phase diagram changes dramatically with interaction strength (Figs. \ref{phase_diagram_weak_interaction} and \ref{phase_diagram_stronger_interactions}). These phase diagrams, and even the equation of state, can be probed in experiments \cite{3DsmallFFLO+BP1indeepBEC+detection+noncond, 3DsmallFFLO+BPintro+BP1indeepBEC+detection, 1drichphases+detection, quasi1Dpromising+detection, BPintro+detection, quasi1Dpromising+quasi1Dsingleband+detection, DFS+detection, mixedphase+detection, quasi1Dpromising+bolechFFLOdetection, detection+, detection}.

\section{\label{model}Model}
Our starting point is the many-body Hamiltonian
\begin{align}
\nonumber \hat{H} = \int d^3 r \Big[&\sum_{\sigma=\uparrow,\downarrow} \hat{\psi}_{\sigma}^{\dagger}(\vec{r}) \big(\hat{H}^{\mbox{\scriptsize{sp}}} - \mu_{\sigma} \big) \hat{\psi}_{\sigma} (\vec{r}) \\
& + g \;\hat{\psi}_{\uparrow}^{\dagger} (\vec{r}) \hat{\psi}_{\downarrow}^{\dagger} (\vec{r}) \hat{\psi}_{\downarrow} (\vec{r}) \hat{\psi}_{\uparrow} (\vec{r}) \Big]\;,
\end{align}
where $\hat{\psi}_{\sigma} (\vec{r})$ denote the fermion field operators, $\hat{H}^{\mbox{\scriptsize{sp}}}$ is the single-particle Hamiltonian, $\hat{H}^{\mbox{\scriptsize{sp}}} = -\hbar^2 \nabla^2/(2m) + (1/2) m \omega_{\perp}^2 (x^2 + y^2)$, and $g$ is the `bare' coupling constant describing interactions between an $\uparrow$-spin and a $\downarrow$-spin. We can relate $g$ to $a_s$ by the Lippmann-Schwinger equation $1/g \hspace{-0.05cm}=\hspace{-0.05cm} m/(4\pi \hbar^2 a_s) \hspace{-0.05cm}-\hspace{-0.05cm} \int \hspace{-0.05cm} d^3 k \hspace{0.05cm} m/(8 \pi^3 \hbar^2 k^2)$ \cite{LippmannSchwinger}. We define the pairing field $\Delta(\vec{r}) \hspace{-0.05cm}=\hspace{-0.05cm} g \langle \hat{\psi}_{\downarrow}(\vec{r}) \hat{\psi}_{\uparrow} (\vec{r})\rangle$, and ignore quadratic fluctuations, arriving at the mean-field Hamiltonian
\begin{align}
\nonumber \hat{H}^{\mbox{\tiny{MF}}} = & \int d^3 r 
\left( \begin{matrix} 
\hat{\psi}_{\uparrow} (\vec{r}) \\
\hat{\psi}_{\downarrow}^{\dagger} (\vec{r}) \end{matrix} \right)^{\dagger} 
\left( \begin{matrix}
\hat{H}^{\mbox{\scriptsize{sp}}} - \mu_{\uparrow} & \Delta(\vec{r}) \\
\Delta^{*} (\vec{r}) & \mu_{\downarrow} - \hat{H}^{\mbox{\scriptsize{sp}}}
\end{matrix} \right)
\left( \begin{matrix} 
\hat{\psi}_{\uparrow} (\vec{r}) \\
\hat{\psi}_{\downarrow}^{\dagger} (\vec{r}) \end{matrix} \right) \\
& + \sum_{n} \big( \varepsilon_{n}^{\mbox{\scriptsize{sp}}} - \mu_{\downarrow} \big) - (1/g) \int d^3 r \; |\Delta(\vec{r})|^2 \;,
\end{align}
where $\varepsilon_{n}^{\mbox{\scriptsize{sp}}}$ denote the single-particle energies. We diagonalize $\hat{H}^{\mbox{\tiny{MF}}}$ by a Bogoliubov transformation, obtaining $\hat{H}^{\mbox{\tiny{MF}}} = \sum_{n} [ (E_n - h) \hat{\gamma}_{n_{\uparrow}}^{\dagger}  \hat{\gamma}_{n_{\uparrow}} + (E_n + h) \hat{\gamma}_{n_{\downarrow}}^{\dagger} \hat{\gamma}_{n_{\downarrow}} + (\varepsilon_{n} - E_n)]$ $-(1/g) \int d^3 r |\Delta(\vec{r})|^2$. Here $\varepsilon_n \equiv \varepsilon_{n}^{\mbox{\scriptsize{sp}}} - \mu$, $\hat{\gamma}_{n_{\uparrow,\downarrow}}$ represent the Bogoliubov quasiparticle annihilation operators, and the eigenvalues $E_n$ ($\geqslant 0$) are determined from
\begin{equation}
\left( \begin{matrix}
\hat{H}^{\mbox{\scriptsize{sp}}} - \mu & \Delta(\vec{r}) \\
\Delta^{*} (\vec{r}) & \mu - \hat{H}^{\mbox{\scriptsize{sp}}}
\end{matrix} \right)
\left( \begin{matrix}
u(\vec{r}) \\ v(\vec{r})
\end{matrix} \right) = E
\left( \begin{matrix}
u(\vec{r}) \\ v(\vec{r})
\end{matrix} \right) .
\label{eigenvalue_equation}
\end{equation}
In the zero-temperature ground state, all quasiparticle states with a negative energy are filled, and others are empty, which yields a total energy
\begin{equation}
\mathcal{E} = \sum_n [\alpha(E_n - h) + \varepsilon_{n} - E_n] - (1/g) \int d^3 r \;|\Delta(\vec{r})|^2 \;,
\label{energy}
\end{equation}
where $\alpha(x) \equiv x$ for $x<0$, and 0 for $x>0$. The ground-state solution is found by minimizing $\mathcal{E}$ as a function of $\Delta(\vec{r})$ for a given $\mu$ and $h$.

To simplify calculations, we take the ansatz $\Delta(\vec{r}) = \Delta_0 \exp{[-(x^2+y^2)/\xi^2]} \exp{(i q z)}$, and minimize Eq. (\ref{energy}) with respect to $\Delta_0$, $\xi$, and $q$. The $\exp{(i q z)}$ factor describes Fulde-Ferrell (FF) pairing at wave-vector $q$. The ansatz (with $q=0$) also encompasses the BCS and the BP phases, and when $\Delta_0=0$ includes the Normal phase. A Larkin-Ovchinnikov (LO) ansatz, in which $\exp{(i q z)}$ is replaced by $\cos{(q z)}$, produces very similar results. Based on prior calculations, one expects that further ansatzes, such as the liquid crystal phases in \cite{liquidcrystal}, will also give similar boundaries. While we label regions of the phase diagram as FFLO, the exact nature of the order is uncertain.

We diagonalize Eq. (\ref{eigenvalue_equation}) by expanding $u(\vec{r})$ and $v(\vec{r})$ in the single-particle states with energies lower than a cut-off $E_c$. We exactly solve this finite-dimensional low-energy sector, and calculate the contribution of higher-energy states perturbatively. We write Eq. (\ref{eigenvalue_equation}) in the bra-ket notation, and express $|v \rangle$ in terms of $|u \rangle$ to obtain $(\hat{H}^{\mbox{\scriptsize{sp}}} - \mu) |u \rangle + \hat{\Delta} (\hat{H}^{\mbox{\scriptsize{sp}}} + E - \mu)^{-1} \hat{\Delta}^{\dagger} |u \rangle = E |u \rangle$. The second term acts as a perturbation, yielding $E_n - \varepsilon_n = \langle n | \hat{\Delta} (\hat{H}^{\mbox{\scriptsize{sp}}} + \varepsilon_{n}^{\mbox{\scriptsize{sp}}} - 2\mu)^{-1} \hat{\Delta}^{\dagger} |n \rangle$, where $|n\rangle$ is the corresponding single-particle state. Using completeness of the single-particle states, we write this as $E_n - \varepsilon_n = \int_0^{\infty} d \tau e^{-2\mu\tau} \langle n| e^{-\hat{H}^{\mbox{\tiny{sp}}} \tau} \hat{\Delta} e^{-\hat{H}^{\mbox{\tiny{sp}}} \tau} \hat{\Delta}^{\dagger} |n \rangle$, which can be expanded in powers of $\varepsilon_n^{-1}$ using the Hadamard lemma. Since $\varepsilon_n$ is large, we only retain the first term, which is $\langle n| \hat{\Delta} \hat{\Delta}^{\dagger} |n \rangle / (2\varepsilon_n)$. Thus we rewrite Eq. (\ref{energy}) as
\begin{equation}
\mathcal{E} = \mathcal{E}_{\mbox{\scriptsize{ex}}} - \sum \langle n| \hat{\Delta} \hat{\Delta}^{\dagger} |n \rangle / (2\varepsilon_n) - g^{-1} \int d^3 r |\Delta(\vec{r})|^2,
\label{energyseparated}
\end{equation}
where $\mathcal{E}_{\mbox{\scriptsize{ex}}}$ denotes the exact-diagonalized part, and the sum is over $n$ with $\varepsilon_n^{\mbox{\scriptsize{sp}}} > E_c$. We take $|n\rangle = |n_x,n_y,k\rangle$, where $(n_x,n_y)$ labels harmonic oscillator states in the $x$-$y$ plane, and $k$ labels plane waves along $z$. Then $\varepsilon_n^{\mbox{\scriptsize{sp}}} \hspace{-0.08cm} = (n_x+n_y+1)\hbar\omega_{\perp} + \hbar^2 k^2/(2m)$, and $\langle n| \hat{\Delta} \hat{\Delta}^{\dagger} |n \rangle = \Delta_0^2 \xi^2 / (4 \pi d_{\perp}^2 \sqrt{n_x n_y})$ for large $n_x$, $n_y$, where $d_{\perp}^2 = \hbar/(m\omega_{\perp})$. Thus the energy per unit length along $z$ is
\begin{equation}
\hspace{-0.1cm} \mathcal{\tilde{E}} \hspace{-0.05cm}=\hspace{-0.04cm} \mathcal{\tilde{E}}_{\mbox{\scriptsize{ex}}} \hspace{-0.02cm}+\hspace{-0.01cm} \tilde{\Delta}_0^2 \tilde{\xi}^2/(4\pi) \big[\tilde{k}_c \big(1 \hspace{-0.03cm} + \hspace{-0.03cm} f\big((1-\tilde{\mu})/\tilde{k}_c^2\big)\big) \hspace{-0.02cm} - \hspace{-0.01cm} \pi/(2\tilde{a}_s) \big],
\label{energyfinal}
\end{equation}
where $\tilde{k}_c \equiv (2(\tilde{E}_c-1))^{1/2}$, $f(x) \equiv \sqrt{2x} \hspace{0.05cm} \tan^{-1} \hspace{-0.04cm} \sqrt{2x}$, and the tildes denote non-dimensionalized quantites, with energies rescaled by $\hbar \omega_{\perp}$ and lengths rescaled by $d_{\perp}$. We perform calculations with $\tilde{E}_c = 10$. We verified that our results are unchanged if $\tilde{E}_c$ is made larger. Our approach to including high-energy modes eliminates the ultraviolet divergence associated with the contact interaction. It is similar to the approach in \cite{3DsmallFFLO+Hartree+regularizationLDA}, where higher modes are included via a local-density approximation. Other regularization schemes have also been successful \cite{otherregularizations}.

\section{\label{results}Results of the full model}
Figure \ref{phase_diagram_weak_interaction}a shows the phase diagram at weak interactions. For small $h$ the ground state is a fully paired BCS state. Increasing $h$ drives a first-order transition to an FFLO or a Normal (N) region. As described earlier, in this weak-coupling limit, the phase boundary is reminiscent of 1D, with a structure that repeats with $\mu$ as various channels open. The FFLO state is most stable when $\mu$ is just above $n \hbar \omega_{\perp}$ for integer $n$. The length $\xi$ over which $\Delta(\vec{r})$ falls off increases with $\mu$. The FFLO wave-vector $q$ grows with $h$. The FFLO-Normal and FFLO-FP transitions are second-order, with the amplitude $\Delta_0 \to 0$ as the boundary is approached.
\begin{figure}[t]
\includegraphics[width=\columnwidth]{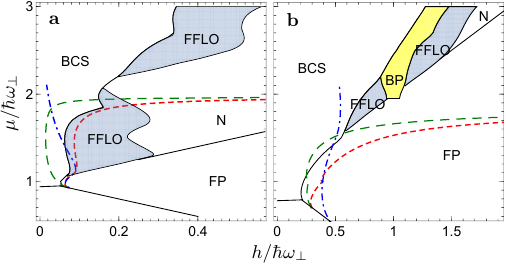}
\caption{\label{phase_diagram_stronger_interactions}(Color online) Zero-temperature phase diagram of the full model for {\bf a}: $d_{\perp}/a_s = -3/2$, {\bf b}: $d_{\perp}/a_s = 0$. Dashed curves plot the BCS critical field predicted by effective 1D models. Conventions for the curves are same as in Fig. \ref{phase_diagram_weak_interaction}.}
\end{figure}
\begin{figure}[t]
\includegraphics[scale=1]{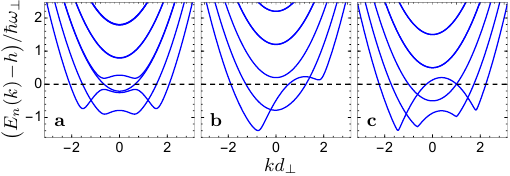}
\caption{\label{quasiparticle_dispersion}(Color online) Quasiparticle dispersion for {\bf a}: the BP phase at $(\tilde{\mu},\tilde{h})=(2.2,1)$, {\bf b}: the FFLO phase at $(\tilde{\mu},\tilde{h})=(2,0.8)$, {\bf c}: the FFLO phase at $(\tilde{\mu},\tilde{h})=(2.35,1.15)$. Different curves denote different transverse modes.}
\end{figure}

Figure \ref{phase_diagram_stronger_interactions} shows how the phase diagram changes at stronger interactions. As interactions favor pairing, we find superfluidity over a larger area. However, the phase diagram becomes more 3D-like, and the relative stabilities of different superfluid phases change. In particular, we see the appearance of a stable breached-pair (BP) phase near unitarity. As seen in the excitation spectra in Fig. \ref{quasiparticle_dispersion}a, the BP state is a gapless superfluid with a uniform order-parameter (in the $z$ direction), which contains both paired and unpaired modes. The unpaired fermions fill the sea of negative energy states. The literature (mostly on isotropic systems) distinguishes between BP states by the topology of the Fermi sea \cite{BPintro+detection, BPintro+quarkmatter+BP1indeepBEC+topology, BP1indeepBEC}. For a given transverse quantum number, the Fermi sea in Fig. \ref{quasiparticle_dispersion}a is connected, making our state analogous to the ``BP1" state in \cite{BPintro+detection}. We do not find BP states where a Fermi sea is broken into disjoint momentum-intervals (cf. \cite{3DsmallFFLO+BP1indeepBEC+detection+noncond, 3DsmallFFLO+BPintro+BP1indeepBEC+detection, 3DsmallFFLO+BPintro+BP1indeepBEC, 3DsmallFFLO+BP1indeepBEC+noncond, 3DsmallFFLO+BP1indeepBEC, BPintro+quarkmatter+BP1indeepBEC+topology, BPintro, quasi1Dfinitesize+BP, BP1indeepBEC, interiorgap, topology, BP+display}). However, we do find FFLO states of both varieties (Fig. \ref{quasiparticle_dispersion}b-c). The BCS-BP transition, as well as the BP-FFLO transition are first-order, accompanied by jumps in the polarization.
\begin{figure}[t]
\includegraphics[width=\columnwidth]{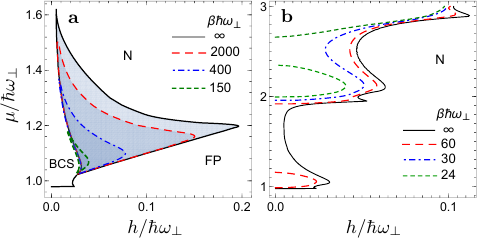}
\caption{\label{temperature_variation1}(Color online) Variation of the superfluid regions with temperature for $d_{\perp}/a_s=-3$. {\bf a}: FFLO region. {\bf b}: BCS region(s). The BCS phase is stable to the left of the curve(s) at a given temperature.}
\end{figure}
\begin{figure}[t]
\includegraphics[width=\columnwidth]{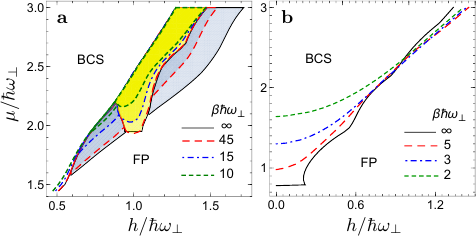}
\caption{\label{temperature_variation2}(Color online) Variation of the superfluid regions with temperature for $d_{\perp}/a_s=0$. {\bf a}: FFLO region. {\bf b}: BCS region.}
\end{figure}

We show the phase diagrams at finite temperature in Figs. \ref{temperature_variation1} and \ref{temperature_variation2}. Here we include thermal fluctuations at temperature $T$ by minimizing the mean-field free energy $F = \mathcal{E} - T S$, where $S$ denotes the entropy. This has the effect of changing the sum in Eq. (\ref{energy}) to $(-1/\beta) \sum_n \ln (1+e^{-\beta(E_n-h)}) + \sum_n (\varepsilon_n+h)$, where $\beta \equiv 1/(k_B T)$, and $E_n$ takes on both positive and negative values. Such a mean-field approach ignores the contribution of non-condensed pairs, and overestimates the critical temperature \cite{3DsmallFFLO+BP1indeepBEC+detection+noncond,3DsmallFFLO+BP1indeepBEC+noncond,fluctuations+noncond, noncond+,noncond}. However, we expect the qualitative features in Figs. \ref{temperature_variation1} and \ref{temperature_variation2} to be valid. In particular, we find vastly different critical temperatures for the FFLO and BCS phases, requiring separate figures to show the behavior. This separation of scales is reasonable, as the pairing energy of the gapped BCS phase is much larger than the gapless FFLO or BP phases.  The critical temperatures grow with the interaction strength since the pairing energy is increased. The BCS phase acquires polarization at finite $T$, which causes $\Delta_0$ to decrease with $h$, making the BCS-Normal transition second-order at small $\tilde{\mu}$. At sufficiently high temperature the BP and BCS phases merge and become indistinguishable. The most striking feature of the weak-coupling phase diagram (Fig. \ref{temperature_variation1}) is that the BCS region breaks up into a series of disconnected lobes which disappear one by one at higher temperatures.

\section{\label{effective1D}Derivation and comparison with an effective 1D model}
To further understand this system, we take $q=0$ and map it onto an effective 1D model for $\tilde{\mu}<2$. We project Eq. (\ref{eigenvalue_equation}) into the harmonic oscillator basis, treating $\Delta_{\vec{m},\vec{n}} \equiv \langle \vec{m}| \hat{\Delta} | \vec{n} \rangle$ as a perturbation if $\vec{n}$ or $\vec{m} \neq \vec{0}$ (where $\vec{n} \equiv (n_x,n_y)$). This yields a 1D BdG equation for the $\vec{n}=\vec{0}$ mode. Neglecting the influence of higher modes on the lowest mode yields an energy per unit length
\begin{align}
\nonumber \tilde{\mathcal{E}} =& \hspace{0.05cm} \frac{\tilde{\Delta}^2 \tilde{\xi}^2}{16 \pi^2} \int \hspace{-0.05cm} \frac{d^3 \tilde{k}}{\tilde{k}^2} - \sideset{}{'}\sum_{\vec{m},\vec{n}} \tilde{\Delta}_{\vec{m},\vec{n}}^2 \int \hspace{-0.05cm} \frac{d \tilde{k}}{4\pi} \frac{\tilde{\varepsilon}_{\vec{m}}/\tilde{\varepsilon}_{\vec{m},+} + \tilde{\varepsilon}_{\vec{n}}/\tilde{\varepsilon}_{\vec{n},+}}{\tilde{\varepsilon}_{\vec{m},+} + \tilde{\varepsilon}_{\vec{n},+}} \\
&-\frac{\tilde{\Delta}^2 \tilde{\xi}^2}{8 \tilde{a}_s} + \hspace{-0.05cm} \int \hspace{-0.05cm} \frac{d \tilde{k}}{2\pi} \big[ \tilde{\varepsilon}_{\vec{0}} - \tilde{\varepsilon}_{\vec{0},+} \hspace{-0.05cm} + \alpha \big(\tilde{\varepsilon}_{\vec{0},+} \hspace{-0.07cm} - \tilde{h}\big) \big],
\label{second_order_energy}
\end{align}
where the integrals are over all $\tilde{k}$, and the prime on the sum stands for $(\vec{m},\vec{n}) \neq (\vec{0},\vec{0})$. Here $\varepsilon_{\vec{0},+} = (\varepsilon_{\vec{0}}^2 + \Delta_{\vec{0},\vec{0}}^2)^{\frac{1}{2}}$, and $\varepsilon_{\vec{n},+} = \varepsilon_{\vec{n}}$ for $\vec{n} \neq \vec{0}$, with $\tilde{\varepsilon}_{\vec{n}} = n_x + n_y + \tilde{k}^2/2+1-\tilde{\mu}$. The first two terms in Eq. (\ref{second_order_energy}) separately diverge, but their sum is finite. This expression for $\tilde{\mathcal{E}}$ maps to that of a purely 1D mean-field model provided we identify the effective 1D order parameter $\Delta_{\mbox{\scriptsize{1D}}}$ and the coupling constant $g_{\mbox{\scriptsize{1D}}}$ as $\Delta_{\mbox{\scriptsize{1D}}} = \Delta_{\vec{0},\vec{0}} =\Delta_0 \tilde{\xi}^2/(\tilde{\xi}^2+1)$, and
\begin{align}
\nonumber \frac{1}{\tilde{g}_{\mbox{\scriptsize{1D}}}} &=\hspace{0.05cm} \frac{(\tilde{\xi}^2+1)^2}{8\tilde{\xi}^2 a_s} - \lim_{n_c \to \infty} \bigg[ \frac{(\tilde{\xi}^2+1)^2}{8\tilde{\xi}^2} \sqrt{2 n_c - 2\tilde{\mu} + 3} \\
&- \sideset{}{'}\sum_{\vec{m},\vec{n}} \mathcal{C}_{m_x n_x} \mathcal{C}_{m_y n_y} \int \hspace{-0.07cm} \frac{d \tilde{k}}{4\pi} \frac{\tilde{\varepsilon}_{\vec{m}}/\tilde{\varepsilon}_{\vec{m},+} + \tilde{\varepsilon}_{\vec{n}}/\tilde{\varepsilon}_{\vec{n},+}}{\tilde{\varepsilon}_{\vec{m},+} + \tilde{\varepsilon}_{\vec{n},+}} \bigg].
\label{g1d}
\end{align}
Here $\tilde{g}_{\mbox{\scriptsize{1D}}} \equiv g_{\mbox{\scriptsize{1D}}} / (d_{\perp} \hbar\omega_{\perp})$, and $\mathcal{C}_{m n} \equiv (2/(\tilde{\xi}^2+1))^{m+n}$ $\big(\Gamma(\frac{m+n+1}{2})\hspace{0.05cm} _2F_1(-m,-n;\frac{1-m-n}{2}; \frac{\tilde{\xi}^2+1}{2})\big)^2 /(\pi m! n!)$ when $m+n$ is even, and 0 otherwise, $_2F_1$ being a hypergeometric function. The prime on the sum now stands for $2 \leqslant m_x+n_x+m_y+n_y \leqslant 2 n_c$. The expression in Eq. (\ref{g1d}) converges as $n_c^{-3/2}$. %When $q \neq 0$, $\tilde{\mu}$ gets replaced by $\tilde{\mu} - \tilde{q}^2/8$ in Eq. (\ref{g1d}).
The effective coupling constant $g_{\mbox{\scriptsize{1D}}}$ is weakly dependent on $\Delta(\vec{r})$, and its structure is best understood by taking $\tilde{\Delta}_0 \to 0$, $\tilde{\xi} \to 1$, for which
\begin{align}
\nonumber & \frac{1}{\tilde{g}_{\mbox{\scriptsize{1D}}}} = \frac{1}{2\tilde{a}_s} + \frac{\zeta(\frac{1}{2},2-\tilde{\mu})}{2\sqrt{2}} \\
&- \frac{\sqrt{2}}{\pi} \Theta(\tilde{\mu}-1) \sum_{j=1}^{\infty} \frac{2^{-2j}}{\sqrt{j \hspace{-0.05cm} + \hspace{-0.05cm} 1-\tilde{\mu}}} \tan^{-1} \sqrt{\frac{\tilde{\mu}-1}{j \hspace{-0.05cm} + \hspace{-0.05cm} 1 -\tilde{\mu}}} \hspace{0.1cm},
\label{mapping}
\end{align}
where $\zeta$ denotes the Hurwitz zeta function, and $\Theta$ is the unit step function. At $\tilde{\mu} = 1$, $1/\tilde{g}_{\mbox{\scriptsize{1D}}} = 1/(2\tilde{a}_s) + \zeta(1/2)/(2\sqrt{2})$, which is Olshanii's two-particle result \cite{mapping}. As $\tilde{\mu}$ grows, $\tilde{g}_{\mbox{\scriptsize{1D}}}$ decreases, approaching $-\infty$ as $\tilde{\mu} \to 2$. This divergence is unphysical, and signals a breakdown of the mapping to 1D when more channels open.

In Fig. \ref{phase_diagram_weak_interaction}b we evaluate the validity of this mapping by plotting the critical field of the BCS phase, $h_c$, from the effective 1D model. It closely follows the critical field obtained from the full model for nearly all $\tilde{\mu}<2$. We also plot $h_c$ using Olshanii's mapping \cite{mapping}, which agrees with the full model at small $\tilde{\mu}$, but becomes less accurate as $\tilde{\mu}$ increases. Further, we show the prediction of the Bethe Ansatz with the mapping in Eq. (\ref{mapping}), which illustrates the difference between an exact and a mean-field analysis in 1D \cite{1DrichphasesMF+mapping+1Dphasediagram+exactvsMF}. The mapping to 1D becomes less accurate at stronger interactions due to mixing of the trap levels, as seen in Fig. \ref{phase_diagram_stronger_interactions}.

\section{\label{outlook}Outlook}
Achieving the temperatures required to directly observe the FFLO state at weak coupling is extremely challenging. The numbers near unitarity are more promising, but the accuracy of our mean-field theory is questionable there. The 1D thermodynamic measurements are promising \cite{1Drichphases+1Dexperiment+1Dphasediagram}. Time-dependent BdG calculations suggest the FFLO domain walls will be observable in time-of-flight expansion of 1D gases \cite{quasi1Dpromising+bolechFFLOdetection}. This signature should be even more robust in the geometries we have been studying. There are also interesting connections to experiments on domain walls in highly elongated traps \cite{solitonvortex}. It is likely that these various research directions will converge in the near future.

\section{\label{acknowledgments}Acknowledgments}
We thank Randy Hulet and Ben Olsen for useful discussions. This work was supported by the National Science Foundation under Grant PHY-1508300.

\end{document}